\documentclass[12pt,a4paper]{article}

\usepackage{graphicx}        

\title{On a flow of substance in a channel of network that contains a main arm and two branches}

\author{Nikolay K. Vitanov, Roumen Borisov}

\date{Institute of Mechanics, Bulgarian Academy of Sciences, Acad. G. Bonchev Str., Bl. 4, 1113 Sofia, Bulgaria}
\begin{document}
\maketitle

\abstract{We study  the motion of a substance in a channel that is part of a network. The channel has 3 arms and consists of  nodes of the network and edges that connect the nodes and form  ways for motion of the substance. Stationary regime of the flow of the substance in the  channel is discussed and statistical distributions for the amount of substance in the nodes of the channel are obtained. These distributions for each of the three arms of the channel contain as particular case famous long-tail distributions such as Waring distribution, Yule-Simon distribution and Zipf distribution. The obtained results are discussed from the point of view of technological applications of the model (e.g., the motion of the substance is considered to happen in a complex technological system and the obtained analytical relationships for the distribution of the substance in the nodes of the channel represents the distribution of the substance in the corresponding cells of the technological chains). A possible application of the obtained results for description of human migration in migration channels is  discussed too.
}

\section{Introduction}
The studies on the  dynamics of complex systems are very intensive in the last decade 
especially in the area of nonlinear dynamics \cite{a1}, \cite{a2}, \cite{dx1}-\cite{d4}, \cite{e1},\cite{k1}, \cite{p1},\cite{s1},\cite{vx1} - \cite{vx6}
and in the areas of social dynamics and population dynamics \cite{albert}, \cite{amaral}, \cite{bok},\cite{marsan},\cite{sat},\cite{elena1},\cite{vk1} - \cite{va15}. 
Networks and the flows in networks are important part of the structure and processes of many complex 
systems \cite{dorog,ff,lu}. Research on network flows  has   roots in the studies on transportation  problems \cite{ff} and in the studies on migration flows \cite{mf2,mf1,v1}. Today one uses the methodology from 
the theory of network flows \cite{ch1} to solve problems connected to e.g., minimal cost of the flow, just in time scheduling or electronic route guidance 
in urban traffic networks \cite{hani}. 
\begin{figure}[!htb]
\centering
\includegraphics[scale=.6]{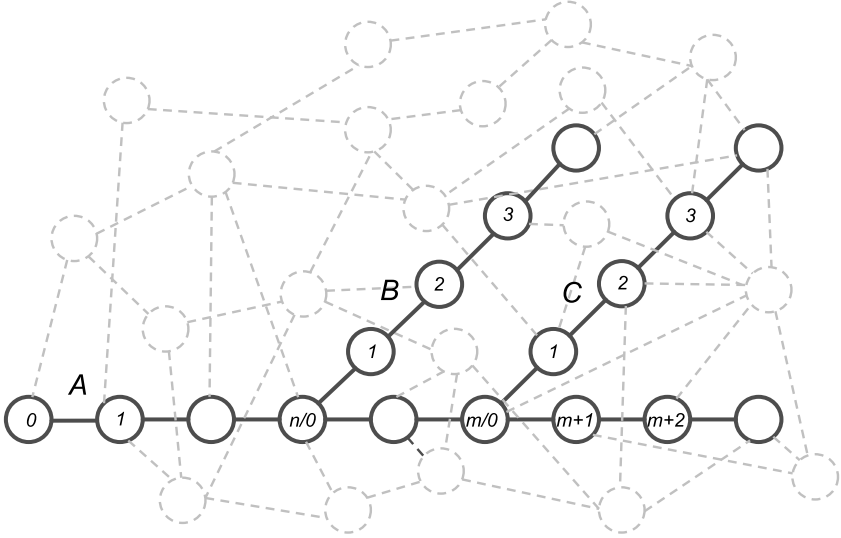}
\caption{Part of a network and a channel consisting of three arms - $A, B$ and $C$. The nodes of the channel are represented by circles and the edges that connect the nodes are represented by solid lines. The main arm ($A$) splits into two arms at two of the nodes. These nodes are labeled $n/0$ and $m/0$ respectively ($m>n$). The entry nodes of the channels $A, B$ and $C$ are labeled by $0$. The  nodes of the network that are not a part of the studied channel are connected by edges that are represented by dashed lines. The assumption below is that the lengths of the arms of the channel are infinite.}
	\label{fig:net2}
\end{figure}
A specific feature of the study below is that we consider a  channel of a network and this
channel consists of three arms. Several types of channels are possible depending on the 
mutual arrangement of the arms. In this study we shall consider the situation where the 
channel consists a main arm ($A$, Fig. \ref{fig:net2}) and two other arms connected 
to the main one in two different nodes ($n/0$ and $m/0,\ m>n$). In this way the channel 
has a single arm up to the node $n/0$ where the channel splits to two arms ($A$ and $B$). The first 
arm  $A$ continues up to the node $m/0$ where it also splits to two arms - $A$ and $C$. Thus the studied channel has three arms: $A$, $B$, and $C$.
\par
There are three special nodes in this channel. The first node of the channel arm $A$ (called also the entry node and labeled by the number $0$) is the only node of the channel where the substance may enter the channel from the
outside world. After that the substance flows along the channel $A$ up to the node labeled by $n/0$. This is the second special node where the channel splits for the first time and this node (node $n/0$)   is the entry node for the second arm $B$ of the channel. The third special node of the network is the node where  the channel splits in two arms for the second time (node $m/0$). This node is the entry node for the third arm $C$ of the channel. We assume that the substance moves only in one direction along the channel (from nodes labeled by smaller numbers to nodes labeled by larger numbers). In addition we assume that the substance may quit the channel and may move to the environment. This
process will be denoted below as "leakage". As the substance can enter the network only through the entry node of the channel arm $A$ then the "leakage" is possible only in the direction from the channel to the  network (and not in the opposite direction). 
\section{Mathematical formulation of the problem}
As we have mentioned above the studied channel consists of three arms.
The nodes on each arm are connected by edges and each node is connected only to the two neighboring nodes of the arm exclusive for the first  node of the channel arm $A$ that is connected only to the neighboring node and the two other special nodes that are connected to three nodes each. We consider each node as a cell (box), i.e.,  we consider an  array of infinite  number of cells indexed in succession by non-negative integers. We assume that an amount $x^q$ of some substance  is distributed among the cells of the arm $q$ ($q$ can be $A$, $B$ or $C$) and this substance can move from one cell to the neighboring cell. 
\par
Let $x_i^q$ be the amount of the substance in the $i$-th cell on the $q$-th arm. We shall consider channel containing infinite number of nodes in each of its three arms. Then
the substance in the $q$-th arm of the channel is
\begin{equation}\label{eq1}
	x^q = \sum \limits_{i=0}^\infty x_ i^q, \ \ q = \{A,B,C\}.
\end{equation}
The fractions $y_i^q = x_i^q/x^q$ can be considered as probability values of distribution of a discrete random variable $\zeta$ in the corresponding arm of
the channel
\begin{equation}\label{eq2}
	y_i^q = p^q(\zeta = i), \ i=0,1, \dots
\end{equation}
The content $x_i^q$ of any cell may change due to the following 4 processes:
\begin{enumerate}
	\item Some amount $s^A$ of the substance $x^A$  enters the main arm $A$ from the external environment through the $0$-th cell; 
	\item Some amount $s^q$ of the substance $x^q$  enters the $q$-th arm from the main one through the $n/0$-th or $m/0$-th cell, respectively ($q \in \{B,C\}$);
	\item Rate $f_i^q$ from $x_i^q$ is transferred from the $i$-th cell into the $i+1$-th cell of the $q$-th arm;
	\item Rate $g_i^q$ from  $x_i^q$  leaks out the $i$-th cell of the $q$-th arm into the environment;
\end{enumerate}
We assume that the process of the motion of the substance is continuous in the time. Then the process can be modeled mathematically by the system of ordinary differential equations:
\begin{eqnarray} \label{eq3}
	\frac{dx_0^q}{dt} &=& s^q-f_0^q-g_0^q; \nonumber \\
	\frac{dx_i^q}{dt} &=& f_{i-1}^q -f_i^q - g_i^q, \ i=1,2,\dots,;q=\{A,B,C\}
\end{eqnarray}
\par
There are  two regimes of functioning of the channel: stationary regime  and non-stationary regime. What we shall discuss below is the stationary regime of functioning of the channel. In the stationary regime of the functioning of the channel $dx_i^q/dt=0$, $i=0,1,\dots$. Let us mark the quantities for the stationary case with $^*$. Then from Eqs.(\ref{eq3}) one obtains 
\begin{equation}\label{eq4}
	f_0^{*q}=s^{*q}-g_0^{*q}; \ \  f_i^{*q}=f_{i-1}^{*q}-g_i^{*q}.
\end{equation}  
This result can be written also as
\begin{equation}\label{eq5}
	f_i^{*q} = s^{*q}- \sum \limits_{j=0}^i g_j^{*q}
\end{equation}
Hence for the stationary case the situation in the each arm is determined by the quantities $s^{*q}$ and $g_j^{*q}$, $j=0,1,\dots$. In this chapter we shall assume the following forms of rules for the motion of substance in Eqs.(\ref{eq3}) ($\alpha_i, \beta_i, \gamma_i, \sigma$ are constants)
\begin{eqnarray}\label{eq6}
	s^A &=& \sigma x_0^A  > 0  \nonumber \\
	s^B&=& \delta_n^A x_n^A; \ \ s^C= \delta_m^A x_m^A ; \ \ 1 \geq \delta^A_{m,n}\geq 0 \nonumber \\
	f_i^q &=& (\alpha_i^q + \beta_i^q i) x_i^q; \ \ \ 1 > \alpha_i^q >0, \ 1 \geq \beta_i^q \ge 0  \nonumber \\
	g_i^q &=& \gamma^{*q}_i x_i^q; \ \ \ 1 \geq \gamma_i^{*q}\ge 0 \to \textrm{non-uniform leakage 
		in the nodes}
\end{eqnarray}
$\gamma^*_i$ is a quantity specific for the present study. ${\gamma_i^*}^A = \gamma_i^A + \delta_i $ describes the situation with the leakages in the cells on the main arm $A$. We shall assume that  $\delta_i^B=0$, $\delta_i^C=0$ for all $i$ and $\delta_i^A=0$ for all $i$ except for $i=n$ and $i=m$. This means that in the $n$-th node in addition to the usual leakage $\gamma_n^A$ there is additional leakage of substance  given by the term $\delta_n^A x_n^A$ and this additional leakage supplies the substance that then begins its motion along the second arm of the channel. Furthermore  in the $m$-th node (where the main arm of the channel splits for the second time) in addition to the usual leakage $\gamma_m^A$ there is additional leakage of substance  given by the term $\delta_m^A x_m^A$ and this additional leakage supplies the substance that then begins its motion along the third arm of the channel.
\par
On the basis of all above the model system of differential equations for each arm of the channel becomes
\begin{eqnarray} \label{eq7}
	\frac{dx_{0}^q}{dt}&=&s^q-\alpha_0^q x_0^q-\gamma_0^{*q} x_0^q ; \nonumber \\
	\frac{dx_{i}^q}{dt}&=&[\alpha_{i-1}^q+(i-1)\beta_{i-1}^q]x_{i-1}^q-(\alpha_{i}^q+ i \beta_{i}^q+\gamma_{i}^{*q})x_{i}^q;\ \ i=1,2,\dots 
\end{eqnarray}
\par
Below we shall discuss the situation in which the stationary state is established in the entire channel. Then $dx_0^q/dt=0$ from the first of the Eqs.(\ref{eq7}). Hence 
\begin{equation}\label{eq7a}
	x_0^q=\frac{s^q}{\alpha_0^q+{\gamma_0^*}^q} \ \ .
\end{equation}
For the main arm  $A$ it follows that $\sigma = \alpha_0^A + \gamma_0^A$. This means that $x_0^A$ (the amount of the substance in the $0$-th cell of the arm $A$) is free parameter. For the arm $B$ and $C$,  $dx_0^k/dt=0$ follows that 
\begin{equation}\label{eq7b}
	x_0^{*B} = \frac{\delta_n^A x_n^{*A}}{\alpha_0^B + \gamma_0^B};\ \ x_0^{*C} = \frac{\delta_m^A x_m^{*A}}{\alpha_0^C + \gamma_0^C}.
\end{equation}
\par
The solution of Eqs.(\ref{eq7}) is (see Appendix A)
\begin{eqnarray}\label{eq8}
	x_i^A = x_i^{*A} &+& \sum \limits_{j=0}^i b_{ij}^A \exp[-(\alpha_j^A + j \beta_j^A + \gamma_j^{*A})t];\ \ i=1,2,\dots\nonumber \\
	x_i^q = x^{*q}_i &+& \sum_{j=0}^{i} b^{q}_{ij}\exp[-(\alpha_j^q + j \beta_j^q + \gamma_j^{*q})t],
	i=1,2,\dots; q=\{B,C\}  \nonumber \\
\end{eqnarray}
where $x_i^{*q}$ is the stationary part of the solution. For $x_i^{*q}$ one obtains the relationship (just set $dx^q/dt = 0$ in the second of Eqs.(\ref{eq7}))
\begin{equation}\label{eq9}
	x_i^{*q} = \frac{\alpha_{i-1}^q + (i-1) \beta_{i-1}^q}{\alpha_i^q + i \beta_i^q + \gamma_i^{*q}} x_{i-1}^{*q}, \ i=1,2,\dots; q=\{A,B,C\}
\end{equation}
\par
The corresponding relationships for the coefficients $b_{ij}^q$ are ($i=1,\dots,q=\{A,B,C\}$):
\begin{equation}\label{eq10}
	b_{ij}^q = \frac{\alpha_{i-1}^q + (i-1) \beta_{i-1}^q}{(\alpha_i^q - \alpha_j^q) + (i \beta_i^q -
		j \beta_j^q) + (\gamma_i^{*q} - \gamma_j^{*q})} b_{i-1,j}^q,
	\ j=0,1,\dots,i-1,
\end{equation}
\begin{figure}[!htb]
\centering
\includegraphics[ scale=0.6]{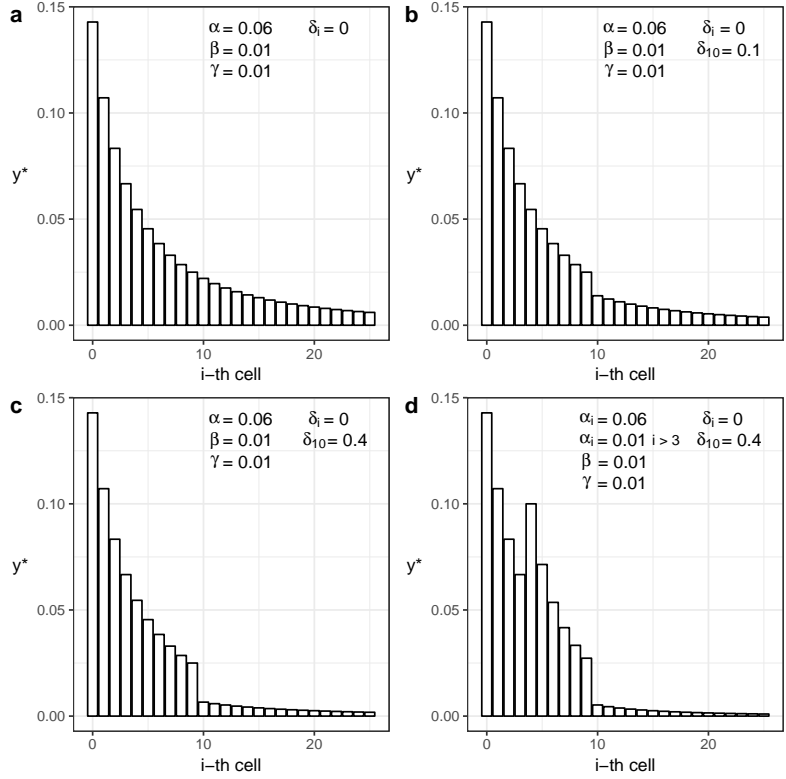}
\caption{The distribution of the substance in the main arm of the channel described by the Eq.(\ref{eq12}) for various values of the  parameters $\gamma_i^*$. Values of the parameters $\alpha_i=0.08,\ \beta_i=0.01$ and $\gamma_i=0.01$ are fixed. The figures show the influence of different choices of the parameter $\gamma_i^* (\delta_i)$ on the shape of the distribution. Figure (3a): $ \delta_i^A=0$. There is a flow of substance only in the main arm $A$ of the channel.  Figure (3b):  $\delta_i^A=0, \ \delta_5^A=0.01$.  In addition to the main arm, there is a flow of substance in the second arm $B$ of the channel too. Figure~(3c):  $\delta_i^A=0,\ \delta_5^A=0.1, \ \delta_{10}^A=0.1$. In this case the substance flows in all three arms of the channel. In addition to the main channel, the second and third channels are also present.  Figure (3d):   $\delta_i=0,\ \delta_5=0.1, \ \delta_{10}=0.4$. }
	\label{fig:2}
\end{figure}
From Eq.(\ref{eq9}) one obtains
\begin{equation}\label{eq11}
	{x_i^q}^* = \frac{\prod \limits_{j=0}^{i-1}[\alpha_{i-j-1}^q+(i-j-1)\beta_{i-j-1}^q]}{
		\prod \limits_{j=0}^{i-1} \alpha_{i-j}^q + (i-j) \beta_{i-j}^q + {\gamma^q}^*_{i-j}} {x_0^q}^*
\end{equation}
The form of the corresponding stationary distribution ${y_i^q}^* = {x_i^q}^*/{x^q}^*$  (where ${x^q}^*$ is the amount of the substance in all of the cells of the arm of the channel) is
\begin{equation}\label{eq12}
	{y_i^q}^* = \frac{\prod \limits_{j=0}^{i-1}[\alpha_{i-j-1}^q+(i-j-1)\beta_{i-j-1}^q]}{
		\prod \limits_{j=0}^{i-1} [\alpha_{i-j}^q + (i-j) \beta_{i-j}^q + {\gamma^q}^*_{i-j}]} {y_0^q}^*
\end{equation}
To the best of our knowledge the distribution presented by Eq.(\ref{eq12}) was not discussed up to now outside our research group, i.e. this is a new statistical distribution.
Let us write the values of this distribution for the first 5 nodes of the main arm of the channel. We assume that the arm $B$ splits at the 3rd node of the arm $A$ and the arm $C$ splits at the 5th node of the arm $A$. Thus the values of the distribution for the first 5 nodes of the channel $A$ are

\begin{eqnarray}\label{valdis}
y_1^{A^*}&=& y_0^{A^*} \frac{\alpha_0^A}{\alpha_1^A + \beta_1^A + \gamma_1^A} \nonumber \\
y_2^{A^*}&=& y_0^{A^*} \frac{\alpha_0^A}{\alpha_1^A + \beta_1^A + \gamma_1^A} \frac{\alpha_1^A + \beta_1^A}{\alpha_2^A + 2 \beta_2^A + \gamma_2^A} \nonumber \\
y_3^{A^*}&=& y_0^{A^*} \frac{\alpha_0^A}{\alpha_1^A + \beta_1^A + \gamma_1^A} \frac{\alpha_1^A + \beta_1^A}{\alpha_2^A + 2 \beta_2^A + \gamma_2^A}
\frac{\alpha_2^A + 2 \beta_2^A}{\alpha_3^A + 3 \beta_3^A + \gamma_3^A + \delta_3} \nonumber \\
y_4^{A^*}&=& y_0^{A^*} \frac{\alpha_0^A}{\alpha_1^A + \beta_1^A + \gamma_1^A} \frac{\alpha_1^A + \beta_1^A}{\alpha_2^A + 2 \beta_2^A + \gamma_2^A}
\frac{\alpha_2^A + 2 \beta_2^A}{\alpha_3^A + 3 \beta_3^A + \gamma_3^A + \delta_3} 
\frac{\alpha_3^A + 3 \beta_3^A}{\alpha_4^A + 4 \beta_4^A + \gamma_4^A}\nonumber \\
y_5^{A^*}&=& y_0^{A^*} \frac{\alpha_0^A}{\alpha_1^A + \beta_1^A + \gamma_1^A} \frac{\alpha_1^A + \beta_1^A}{\alpha_2^A + 2 \beta_2^A + \gamma_2^A}
\frac{\alpha_2^A + 2 \beta_2^A}{\alpha_3^A + 3 \beta_3^A + \gamma_3^A + \delta_3} 
\frac{\alpha_3^A + 3 \beta_3^A}{\alpha_4^A + 4 \beta_4^A + \gamma_4^A} \times \nonumber \\
&&
\frac{\alpha_4^A + 4 \beta_3^A}{\alpha_5^A + 5 \beta_4^A + \gamma_5^A + \delta_5}
\nonumber \\
\end{eqnarray}
 Let us show that the distribution described by Eq.(\ref{eq12}) contains as particular cases several famous distributions, e.g., Waring distribution, Zipf distribution, and Yule-Simon distribution. In order to do this we consider the particular case when $\beta_i^q \ne 0$ and write $x_i^q$ from Eq.(\ref{eq11}) 
by means of the new notations $k_i^q = \alpha_i^q/\beta_i^q$; $a_i^q = {\gamma^q}^*_i/\beta_i^q$;  $b_i^q = \beta_{i-1}^q/\beta_i^q$. 
Let us now consider the particular case where $\alpha_i^q = \alpha^q$ and $\beta_i^q = \beta^q$ for $i=0,1,2,\dots$, $q=\{A,B,C\}$. Then from Eq.(\ref{eq12})  one obtains
\begin{equation}\label{eq15}
	{x_i^q}^* = \frac{[k^q+(i-1)]!}{(k^q-1)! \prod \limits_{j=1}^i (k^q+j+a_j)} {x_0^q}^*
\end{equation}
where $k^q = \alpha^q/\beta^q$ and $a_j^q={\gamma^q}^*_j/\beta^q$. 
Let us now consider the particular case where $a_0^q = \dots = a_N^q$. In this case the distribution for $P^q(\zeta=i)$ for ${y_i^q}^* = {x_i^q}^*/{x^q}^*$ is:
\begin{eqnarray}\label{eq17}
	P^q(\zeta = i) &=& P^q(\zeta=0) \frac{(k^q-1)^{[i]}}{(a^q+k^q)^{[i]}}; \ \ {k^q}^{[i]} = \frac{(k^q+i)!}{k^q!}; \ i=1, 2, \dots 
\end{eqnarray}
$P^q(\zeta=0)={y_0^q}^* = {x_0^q}^*/{x^q}^*$ is the percentage of substance that is located in the first cell of the channel. Let this percentage be 
${y_0^q}^* = \frac{a^q}{a^q+k^q}$
This case  corresponds to the situation where the amount of substance in the first cell is proportional of the amount of substance in the entire channel. In this case Eq.(\ref{eq17}) is reduced to:
\begin{eqnarray}\label{eq19}
	P^q(\zeta = i) &=& \frac{a^q}{a^q+k^q} \frac{(k^q-1)^{[i]}}{(a^q+k^q)^{[i]}}; \ \ {k^q}^{[i]} = \frac{(k^q+i)!}{k^q!}; \ i=1, 2, \dots; q=\{A,B,C\} 
	\nonumber \\
\end{eqnarray}
For all $q$ the distribution (\ref{eq19}) is exactly the Waring distribution (probability distribution of non-negative  integers named after Edward Waring - the 6th Lucasian professor of Mathematics in Cambridge from the 18th century)  \cite{varyu1,varyu2}. Waring distribution has the form
\begin{equation}\label{eq20}
	p_l = \rho \frac{\alpha_{(l)}}{(\rho + \alpha)_{(l+1)}}; \
	\alpha_{(l)} = \alpha (\alpha+1) \dots (\alpha+l-1)
\end{equation}
$\rho$ is called the tail parameter as it controls the tail of the Waring distribution. Waring distribution contains various distributions as particular cases. Let $l \to \infty$ Then the Waring distribution is reduced to  the frequency form of the Zipf distribution \cite{chen}
$p_l \approx \frac{1}{l^{(1+\rho)}}$.
If $\alpha \to 0$ the Waring distribution is reduced to the Yule-Simon distribution \cite{simon} 
$p(\zeta = l \mid \zeta > 0) = \rho B(\rho+1,l)$, 
where $B$ is the beta-function. 
\section{Discussion}
On the basis of the obtained analytical relationships for the distribution of the substance in three arms of the channel we can make numerous conclusions.
Let us consider the distribution in the main channel $A$ described by Eq. (\ref{eq12}) from the last section (we have to set $q=A$). Below we shall omit the index $A$ keeping in the mind that we are discussing the main arm of the channel. Let us denote as $y_i^{*(1)}$ the distribution of the substance in the cells of the main arm of the channel for the case of lack of second and third arms. Let $y_i^{*(2)}$ be the distribution of the substance in the cells of the main arm of the channel in the case of the presence of second and the absence of the third arm of the channel. Finally, let  $y_i^{*(3)}$ be the distribution of the substance in the cells of the main arm of the channel in the case of the presence  of both the second and third arms. From the theory in the previous section one easily obtains the relationship
\begin{eqnarray}\label{eq31a}
	\frac{y_i^{*(1)}}{y_i^{*(r)}} = 
	\prod \limits _{j=0}^{i-1} 
	\frac{\alpha_{i-j} + (i-j) \beta_{i-j} + \gamma^*_{i-j}}{\alpha_{i-j} + (i-j) \beta_{i-j} + \gamma_{i-j}} = \prod \limits _{j=0}^{i-1}
	\left[ 1+ \frac{\delta_{i-j}}{\alpha_{i-j} + (i-j) \beta_{i-j} + \gamma_{i-j}}\right], \nonumber \\
\end{eqnarray}
where $r$ can be $2$ or $3$. 
If there are no second and third arms of the channel then the distribution of the substance in the arm $A$ has a standard form of a long-tail distribution like the distribution shown in Fig. 2a. In the case of presence of
additional arms $B$ and $C$ and if $i < n$ then $\delta_i=0$ and there is no difference between the distribution of the substances in the channel with single arm and in the channel with two arms. The difference arises at the splitting cell $n/0$. As it can be easily calculated for $i\ge n$  and $i<m$ Eq.(\ref{eq31a}) reduces to
\begin{eqnarray}\label{eq32a}
	\frac{y_i^{*(2)}}{y_i^{*(1)}} = \frac{1}{ 
		1+ \frac{\delta_{n}}{\alpha_{n} + n \beta_{n} + \gamma_{n}}}, \ \ m> i \ge n \\ \nonumber
\end{eqnarray}
Eq.(\ref{eq32a}) shows clearly that \emph{the presence of the second arm of the channel changes the distribution of the substance in the main arm of the channel}. The "leakage" of the substance to second arm of the channel  may reduce much the tail of the distribution of substance in the main arm of the channel. This can appear as a kink in the distribution similar to the kink that can be seen at Fig. 2b.  
\par
When  $i\ge m$  Eq.(\ref{eq31a}) reduces to
\begin{eqnarray}\label{eq32b}
	\frac{y_i^{*(3)}}{y_i^{*(1)}} = \left[\frac{1}{ 
		1+ \frac{\delta_{n}}{\alpha_{n} + n \beta_{n} + \gamma_{n}}}\right]
	\left[\frac{1}{ 
		1+ \frac{\delta_{m}}{\alpha_{m} + m \beta_{m} + \gamma_{m}}}\right], \ \ i \ge m\\ \nonumber
\end{eqnarray}
Eq.(\ref{eq32b}) shows that the "leakage" of the substance through both the second and the third arms may reduce even more the tail of the distribution of substance in the main arm than only through the second arm. This reduction becomes extremely strong for large values of $\delta_n$ and $\delta_m$  (Figs. 2c and 2d). Furthermore, it is easily seen that \emph{ the presence of the third arm of the channel does not change the distribution of the substance in the second arm.} We note that the form of the distribution (\ref{eq12}) can be different for different values of the parameters of the distribution. One interesting form of the distribution can be observed in Fig. 2d and this form is different than the conventional form of a long-tail discrete distribution shown in Fig. 2a. Thus at Fig. 2 one can obtain visually a further impression that the distribution (\ref{eq12}) is more general than the Waring distribution.
\section{Concluding remarks}
In this text we have discussed one possible case of motion of substance through a network. Namely our attention was concentrated on the directed motion of substance in a channel of network. Specific feature of the study  is that the channel consists of three arms: main arm and two other arms that split from the main arm. We propose a mathematical model of the motion of the substance through the channel and our interest in this study was concentrated on the stationary regime of the motion of the substance through the arms of the channel. The main outcome of the study is the obtained distributions of substance  along the cells of the channel. These distributions have a very long tail in
the form of the distributions depends on the numerous parameters that regulate the motion of the substance through the channel. Nevertheless we have shown that all of the distributions (i.e. the distributions of the substance along the three arms of the channel) contain as particular case
the long-tail discrete Waring distribution which is famous because of the fact that it contains as particular cases the  Zipf distribution and the Yule-Simon distribution that are much used in the modeling of complex  natural, biological, and social systems.   
\par
The model discussed above can be used for study of motion of substance through cells of technological systems. The model can be applied also for study of other system such as channels of human migration or flows in logistic channels.
Initial version of the model (for the case of channel containing one arm) was applied for modeling dynamics of citations and publications in a group of researchers \cite{sg1}. Let us make several notes on the application in the case of human migration as the study of human migration is an actual research topic
\cite{borj} - \cite{champi}, \cite{hott}, \cite{everet}, \cite{massey}, \cite{puu}, \cite{smn}, \cite{skeldon}, \cite{v1} - \cite{will99}. In this case of application of the theory the nodes are the countries that form the migration channel and the edges represents the ways that connect the countries of the channel.  Eqs.(\ref{eq32a},\ref{eq32b}) show that the additional arms of the channel can be used to decrease the "pressure" of migrants in the direction of the
more preferred countries that are relatively away from the initial countries of the channel. This may be done by appropriate increase of the coefficients $\delta_n$ and $\delta_m$ in Eqs.(\ref{eq32a},\ref{eq32b}).
\par
The research presented above is connected to the actual topic of motion of different types of substances along the
nodes and the edges of various kinds of networks.
We intend to continue this research by study of more
complicated kinds of channels and by use more sophisticated model that accounts for more kinds of processes that may happen in connection with the studied network flows.

\begin{appendix}
	\addcontentsline{toc}{section}{Appendices}
	\renewcommand{\thesubsection}{\Alph{subsection}}
	\subsection {Proof that Eq.(\ref{eq8})  is a solution of Eqs.(\ref{eq7}) for main arm of the channel}
	Let us consider the first equation from Eqs.(\ref{eq7}) for main arm. In this case $i=0$ and  Eqs.(\ref{eq8}) becomes $x_0^A=x_0^{*A}+b_{00}^A\exp[-(\alpha_0^A+~\gamma_0^A)t]$. The substitution of the last relationship in the first of the Eqs.(\ref{eq7}) leads to the relationship
	\begin{equation}\label{eq1a}
		0=(\sigma_0-\alpha_0^A-\gamma_0^{*A})x_0^* +b_{00}^A\sigma_0\exp[-(\alpha_0^A+\gamma_0^{*A})t].
	\end{equation}
	Let us assume $\sigma_0=\alpha_0^A+\gamma_0^{*A}$ and $b_{00}^A=0$. Then Eqs.(\ref{eq8}) describes the solution of the first of Eqs.(\ref{eq7}).
	\par
	Let us now consider Eqs.(\ref{eq7}) for the main arm of the channel for $i=1,2,\dots$ Let us fix $i$ and substitute  the first of Eqs.(\ref{eq8}) in the corresponding equation from  Eqs.(\ref{eq7}). The result is 
	\begin{eqnarray} \label{eq2a}
		\sum \limits_{j=0}^{i-1} \exp[-(\alpha_j^A+j\beta_j^A+\gamma_j^{*A})t]\big\{-b_{ij}^A(\alpha_j^A+j\beta_j^A+\gamma_j^{*A})-\\\nonumber
		-b_{i-1,j}^A[\alpha_{i-1}^A+(i-1)\beta_{i-1}^A]+b_{ij}^A(\alpha_i^A+i\beta_i^A+\gamma_i^{*A})\big\}=0
	\end{eqnarray}
	As it can be seen from  Eq.(\ref{eq10}),   Eq.(\ref{eq2a}) is satisfied. 
\end{appendix}

\end{document}